\documentclass{elsart}
\usepackage{graphicx,harvard,amssymb}
\journal{New Astronomy}

\newcommand{\lsim}{\,\lower2truept\hbox{${<\atop\hbox{\raise4truept\hbox{$\sim$}}}$}\,}
\newcommand{\gsim}{\,\lower2truept\hbox{${>\atop\hbox{\raise4truept\hbox{$\sim$}}}$}\,}

\begin{document}

\begin{frontmatter}

\title{Asteroid detection at millimetric wavelengths with the {\sc Planck}
survey}

\author[Cremonese]{G.~Cremonese},
\author[Marzari]{F.~Marzari},
\author[Burigana]{C.~Burigana},
\author[Maris]{M.~Maris}

\vskip 0.5truecm

\address[Cremonese]{INAF-Osservatorio Astronomico, Vicolo Osservatorio 5, I-35122, Padova,
Italy}
\address[Marzari]{Dipartimento di Fisica, Universit\`a di Padova, Via Marzolo 8,
I-35131, Padova, Italy}
\address[Burigana]{Istituto TeSRE, Consiglio Nazionale delle Ricerche, Via Gobetti
101, I-40129, Bologna, Italy}
\address[Maris]{INAF-Osservatorio
Astronomico, Via G.B.~Tiepolo 11, I-34131, Trieste, Italy}

\footnote{The address to which the proofs have to be sent is: \\
Gabriele Cremonese, Osservatorio Astronomico, Vicolo Osservatorio
5, I-35122, Padova, Italy\\
fax: +39-049-8759840\\
e-mail: cremonese@pd.astro.it}

\newpage
\begin{abstract}

The {\sc Planck} mission, originally devised for cosmological
studies, offers the opportunity to observe Solar System objects at
millimetric and submillimetric wavelengths. We concentrate in
this paper on
the asteroids of the Main Belt, a large class of minor bodies in
the Solar System. At present more that 40000  of these asteroids
have been discovered and their detection rate is rapidly increasing.
We intend to estimate the number of asteroids that can
can be detected during the mission and to evaluate the strength
of their signal. We have rescaled the instrument sensitivities,
calculated by the LFI and HFI teams for sources fixed in the sky,
introducing some degradation factors to properly account for
moving objects. In this way a detection threshold is derived
for asteroidal detection that is related to
the diameter of the asteroid and its geocentric distance. We
have developed a numerical code that models the detection of
asteroids in the LFI and HFI channels
during the mission. This code perfoprms a detailed integration
of the orbits of the asteroids
in the timespan of the mission and identifies those bodies that
fall in the beams of Planck and their signal stenght.
According to our simulations, a total of 397 objects
will be observed by {\sc Planck} and an asteroidal body will
be detected in some beam in 30\% of the total sky scan--circles.
A significant fraction (in the range from $\sim 50$ to 100
objects) of the  397 asteroids
will be
observed with a high S/N ratio.\\
Flux measurements of a large sample of asteroids in the
submillimeter and millimeter range are relevant since they
allow to analyze the
thermal emission and its relation to the surface and regolith
properties. Furthermore, it will be possible to check on a wider
base the two standard thermal models, based on a nonrotating or
rapidly rotating sphere.\\
Our method can also be used to separate Solar System sources from
cosmological sources in the survey. This work is based on {\sc
Planck} LFI activities.

\end{abstract}


\end{frontmatter}

\newpage

\section{Introduction}

The {\sc Planck}~ESA
mission\footnote{http://astro.estec.esa.nl/SA-general/Projects/Planck/}
will perform a high-angular resolution mapping of the microwave
sky over a wide range of frequencies, from 30 to 900 GHz. These
data will have important implications on different fields, from
cosmology and fundamental physics to Galactic and extragalactic
astrophysics. The Solar System astronomy will also benefit from
the Planck mission since it will offer the opportunity to
perform a survey of Solar System objects at millimetric
wavelengths. Planck will be sensitive to the millimetric emissions from
planets and from a significant fraction of the asteroidal
population. In this paper we focus on
the detection of asteroid radio
emissions and on the identification of the asteroidal targets that
will be observed during the mission.
The relevant parameters for
the possible detection  of a minor body by {\sc Planck} are the geocentric
distance and
the diameter.
We will concentrate in this paper on Main Belt asteroids that can be more
easily detected since their orbits are at relatively small
geocentric distances and several objects have diameter
larger than 50~km.\\
Radio observations of asteroids from the Earth were performed by
\citeasnoun{Redman1998} at seven different frequencies ranging
from 150 to 870 GHz. Redman focused on the information that radio
data can give to the thermal models of asteroids. Their set of
targets included 5 asteroids already observed by
\citeasnoun{Altenhoff1994} and two new objects.
\citeasnoun{Altenhoff1994} observed, at 250 GHz, 15 among the
largest asteroids, but only to determine the diameter and to
compare their measurements with other methods.\\
The emissivity obtained at radio frequencies may probe the
temperature of the body at different depths on the surface
allowing to derive precise values of the average temperature.
Moreover, relevant physical data concerning the density of the
regolith layer covering the surface of the asteroids can also be
determined. \\
Radio observations of asteroids by {\sc Planck} offer a unique
opportunity to improve our knowledge on the thermal emission in
the millimetric and submillimetric domain, increasing our
knowledge of the physical nature of the surface layers of the
objects. Considering that radio data provide an estimate of the
density on a scale related to the wavelength of the observations,
the advantage of using Planck will be in the possibility of
gathering data on a large frequency spectrum and then to probe the
temperature and regolith density at different depths below the
surface.\\
Furthermore, the {\sc Planck} data can help to refine the
thermophysical models using also physical studies of the thermal
properties of stony and FeNi meteorites as well as inferences
about surface physical properties derived from asteroid radar
studies. It will be possible to relate the {\sc Planck} data to
the two standard models which are commonly invoked to predict the
thermal radiation from asteroids, considering the
asteroid as a nonrotating sphere or a rapidly rotating sphere. \\
For instance Vesta has been observed on a wide range of
wavelengths, from submillimeter to centimeter, and the ratio of
the observed flux, at each wavelength (from 12$\mu$m to 6cm),
divided by the flux expected from a blackbody at the temperature
of a nonrotating asteroid shows a behavior different from the
standard models \cite{Redman1992}. A successful model for the
thermal emission must explain why the apparent flux remains below
the rapidly rotating sphere value, and much below the nonrotating
sphere value, over the entire submillimeter range. Several
mechanisms can lower the apparent temperature at radio wavelengths
as reflections at the outer surface of the regolith, reduction in
the actual temperature of the deeper layers of the regolith due to
diurnal temperature variations at the surface, or scattering by
grains within the regolith which can reduce the emissivity in a
wavelength dependent fashion. The actual emission from an asteroid
may combine all three effects in differing amounts at each
wavelength, but only very few objects have been observed in the
submillimeter and millimeter range and it is not possible to
identify a typical behavior or to relate the results to different
asteroid taxonomic classes.\\
It is worth to point out that the main belt asteroids may
represent a source of background for the {\sc Planck} mission and,
as a consequence, they should be considered in the definition of
the scientific and technical aspects of the mission. \\
The thermal emission of asteroids, apart from their intrinsic
scientific relevance, has also to be considered as a potential
source of spurious detections in the analysis of Galactic or
extragalactic sources. The precise determination at each epoch of
the eventual presence of an asteroid, with its point emission, in
some of the {\sc Planck} beams is a critical requirement to avoid
systematic errors in mapping the astrophysical sources of
radiation. The high nominal accuracy of the {\sc Planck}
instruments suggests a tight control even for the
contamination from asteroids below the detection threshold.\\
Section~2 is devoted to the discussion of the {\sc Planck}
sensitivity for the detection of moving sources at different
frequency channels. The averaged final {\sc Planck} sensitivities
will be rescaled in order to take into account the positions of
the different antenna beams in the telescope field of view and the
main properties of the {\sc Planck} scanning strategy. These
results are used in Section~3 to determine the threshold to detect
asteroids in the {\sc Planck} data streams, according to their
typical temperature and size. In Section~4, we describe our
numerical code to compute the asteroid transits on the different
{\sc Planck} beams and to estimate their signals.  The results of
this analysis are presented in Section~5. Finally, we discuss the
results and draw our main conclusions in Section~6.

\section{{\sc Planck} performances}

The {\sc Planck}
surveyor will observe the sky at nine frequencies with two instruments
having different angular resolutions and sensitivities: the Low
Frequency Instrument (LFI) that covers the range 30-100 GHz, and
the High Frequency Instrument (HFI) that covers the range 143-857
GHz. In Table~1 we report the relevant parameters for  both LFI
and HFI (see columns 1, 4 and 5). The relative bandwidths of the
LFI and HFI are respectively $\simeq 20\%$ and $\simeq 25\%$
(see column 2 of Table 1). \\
\begin{table}
\caption{Instrumental performances and confusion noise estimates
for {\sc Planck} LFI. The bandwidths are 20\% for the LFI channels
and 25\% for the HFI ones. }
 \begin{center}
 \renewcommand{\arraystretch}{0.5}
\begin{tabular}{cccccccccc}
& & & & & & & & \\
\hline
& & & & & & & & \\
$\nu_{\rm eff}$ & $N_{rad;bol}$ & FWHM & $\sigma_{\rm noise}$ &
$\sigma_{\rm noise}$ & $\sigma_{\rm CMB}$ & $\sigma_{\rm ex.sou.}$
& $\sigma_{\rm Gal}$ &
$\sigma_{\rm noise,MAP}$ \\
& & & & & & & \\
(GHz) & (unpol;pol)  & (arcmin) & ($\mu$K) & (mJy) & (mJy) & (mJy)
&
(mJy) & (mJy) \\
& & & & & & & & \\
\hline
& & & & & & & & \\
30  & 4     & 33.6 & 5.1  & 13.4 & 245 & 60 & 100 & 48 \\
& & & & & & & & \\
44  & 6     & 22.9 & 7.8  & 20.5 & 238 & 45 & 45 & 69 \\
& & & & & & & & \\
70  & 12    & 14.4 & 10.6 & 28.0 & 221 & 30 & 15 & 102 \\
& & & & & & & & \\
100 & 32    & 10.0 & 12.8 & 33.2 & 192 & 20 & 7 & 127  \\
& & & & & & & & \\
\hline
& & & & & & & & \\
100 & 4 ; 0 & 10.7 & 2.9 & 8.7 & 192 & 20 & 7 & 127 \\
& & & & & & & & \\
143 & 3 ; 9 & 8.0  & 3.4 & 11.5 & 149 & 15 & 6 & --  \\
& & & & & & & & \\
217 & 4 ; 8 & 5.5  & 3.1 & 11.5 & 82.6 & 8$-$15 & 5
& --  \\
& & & & & & & & \\
353 & 6 ; 0 & 5.0  & 2.4 & 19.4 & 19.1 & 20$-$35 & 18
& -- \\
& & & & & & & & \\
545 & 0 ; 8 & 5.0  & 2.0 & 38 & 1.5 & 45$-$80 & 62
& -- \\
& & & & & & & & \\
857 & 6 ; 0 & 5.0  & 0.9 & 43 & 0.016 & 100$-$180 &
120 & -- \\
& & & & & & & & \\
\hline
\end{tabular}
 \end{center}
\end{table}

In this work we will refer to the nominal {\sc Planck} sensitivity
per resolution element (a squared pixel with side equal to the
Full Width at Half Maximum (FWHM) of the corresponding beam), as
recently revised by the LFI Consortium ({\sc Planck} Low Frequency
Instrument, Instrument Science Verification Review, October 1999,
LFI Design Report, private reference; see also the LFI proposal,
\citeasnoun{Mandolesi1998}) and as reported in the HFI proposal
\cite{Puget1998}. The sensitivity is related to the measurement of
microwave anisotropy in terms of antenna
temperature and flux\\
The normalized sensitivity per pixel in the sky (i.e., the
sensitivity divided by the averaged sensitivity of the observed
pixels in the sky) for a typical {\sc Planck} beam improves with
the module of the ecliptic latitude up to a ring of optimal
sensitivity which contains a small unobserved sky region.  The
normalized sensitivity averaged over the ecliptic longitudes as a
function of the ecliptic colatitude is close to unit at the
ecliptic colatitude $\theta_e \simeq 50^\circ$ and $130^\circ$ and
it is quite well approximated by the law $\sqrt{ {\rm sin}\,
\theta_e / {\rm sin}\, 50^\circ }$ at least for ecliptic latitudes
between about $-70^\circ$ and $+70^\circ$, relevant for Solar
System objects. The detailed behavior of the sensitivity on
$\theta_e$ is clearly dependent on the choice of the scan angle
$\alpha$, the adopted scanning strategy and the beam position on
the telescope field of view.\\
The Solar System bodies are moving objects and most likely they
have a variable flux depending from their heliocentric and
geocentric distances. It means that their observation and estimate
of their fluxes can not take easily advantage from the twice
coverage of the sky, during the planned period of observation of
14 months, nor from the coadding of data from different feeds if
they look at different sky positions at the same time. Further
details on the {\sc Planck} sensitivity relevant for moving bodies
and variable sources can be found in \cite{Burigana2000} and
\cite{Burigana2001}. For the LFI sensitivities we take into
account the most recent computations of the LFI beam positions of
the telescope field of view \cite{Sandri2001}. The HFI channels
can play a crucial role for the study of the moving bodies of the
Solar System, as their spectra are typically close to blackbodies,
or modified blackbodies, at temperatures significantly higher than
the cosmic microwave background (CMB) or are dominated by dust
emission components or by molecular emission lines, particularly
relevant at the highest
{\sc Planck} frequencies (see, e.g., \citeasnoun{Neufeld2000}).\\
The {\sc Planck} receivers are affected by the $1/f^{\alpha}$
(with $\alpha$ in the range from $\sim 1$ to $\sim 2$) noise that
introduces drifts in the time ordered data identified as stripes
in the final maps \cite{Janssen1996}. The time ordered data have
to be cleaned from these drifts by using appropriate data
reduction methods. Destriping algorithms (see, e.g.,
\citeasnoun{Delabrouille1998}, \citeasnoun{Maino2001}, and
references therein) are able to efficiently remove $1/f^{\alpha}$
noise drifts by working directly in the time domain, fully
preserving the time ordering of the data and allowing to exploit
the cleaned data to properly reveal the signatures from moving and
variable sources. Then it is possible to neglect the drift
contamination in the following discussion. We note also that
destriping algorithms (as well as other classes of reduction
methods) use the comparison between repeated observations at
different times of the same sky positions to remove drifts.
Therefore, cleaning the time ordered data from astrophysical
moving or variable sources by properly flagging the time ordered
data may be useful to improve the destriping codes, according to
the S/N ratios of the contaminating sources (the possibility to
use flagged data have been already implemented in the most recent
version of the LFI destriping code, see
\citeasnoun{Stanghellini2001}). Running this task represents a
minor effort for the data reduction pipeline and it can
be implemented starting from the software developed with this work.\\
In addition to the instrumental noise, the sky itself introduces a
confusion noise given by the Galactic and extragalactic
temperature fluctuations and from the CMB anisotropies, relevant
at different angular scales, $\theta \simeq 180^{\circ}/\ell$,
where $\ell$ is the multipole of the harmonic expansion of the
temperature fluctuation pattern. We report in Table 1 the
fluctuation level of the CMB anisotropy assuming a rms
thermodynamic temperature fluctuation of about 95~$\mu$K as
derived for a standard cosmological flat model, approximately
COBE/DMR normalized with cosmological parameters compatible with
the present constraints on CMB fluctuations at moderate
multipoles, as derived from recent balloon experiments. Of course,
the accurate determination of the CMB confusion
noise at small scales is the {\sc Planck} main goal.\\
Then we report the extragalactic source confusion noise (by taking
into account both radio and infrared sources), as evaluated by
\citeasnoun{Toffolatti1998} and revised by
\citeasnoun{Dezotti1999a} and \citeasnoun{Dezotti1999b}. We
estimate a more precise source confusion noise as
$\sigma_{ex.sou.}=\sqrt{C_l}/{\rm (FWHM/rad)}$ where the angular
power spectrum $C_l$ is the sum of the $C_l$ of radio and far-IR
source fluctuations, as quoted by \citeasnoun{Dezotti1999b} for a
conservative detection limit of 1~Jy (see their Table~6). These
estimates may be then pessimistic. On the other hand, it would be
probably difficult to subtract sources in real time with high
accuracy; in addition, the source clustering, neglected in this
work, may significantly increase the source confusion noise, at
least for some cosmological models, particularly at the highest
{\sc Planck} frequencies \cite{Magliocchetti2001}.\\
The Galactic fluctuation levels reported here are only indicative
and refer to moderate and high Galactic latitudes, as large
variations are present in the sky.\\
The relevant global sensitivity for point source detection/observation
is typically assumed
to be the sum in quadrature of all the sources
of confusion noise, multiplied by a proper constant
(typically in the range from $\sim 2$ to 5).
We report also in the last column of Table 1 the sensitivity of
the MAP experiment\footnote{http://map.gsfc.nasa.gov/html/} ($\sim
35 \mu$K, in terms of thermodynamic temperature, on squared pixels
with 0.3$^\circ$ side at each frequency channel, 22, 30, 40, 60
and 90 GHz), rescaled at the corresponding {\sc Planck} beam size.
In fact, we can argue that after the MAP mission the sum of the
CMB and foreground temperature fluctuations on each sky pixel will
be known with the MAP accuracy.\\
CMB anisotropy experiments provide time ordered data and then
maps of temperature fluctuations, $\delta T$,
from the average temperature; the temperature fluctuations
can be expressed in terms of fluctuations
of antenna temperature, $\delta T_{ant}$,
or in terms of fluctuations of thermodynamic temperature,
$\delta T_{therm}$.
Their ratio is given by:

\begin{equation}
{\delta T_{ant} \over \delta T_{therm}} =
{x^2 {\rm exp}(x) \over [{\rm exp}(x)-1]^2 } \, ,
\end{equation}

where $x=h\nu/kT_0$, $T_0 \simeq 2.725$~K being the CMB monopole
thermodynamic temperature as established by \citeasnoun{Mather1999}.
The same formula holds for ratio between
the rms noise of the observed antenna temperature fluctuation,
$\Delta \delta T_{ant}$, and the rms noise of the observed
thermodynamic temperature fluctuation, $\Delta \delta T_{therm}$.\\
In the study of discrete sources it is preferable to express the
signal fluctuations in terms of flux fluctuations. For a point
source with flux $F_{\nu}$ observed with a beam response $J$
normalized to the beam maximum response, the observed antenna
temperature is given in general by:

\begin{equation}
T_{\nu,ant,obs} = 10^{-41} (c^2/2k) [(F_{\nu}/{\rm Jy}) /
(\nu/{\rm GHz})^2] (J/\int_{4\pi} J d\Omega) \, .
\end{equation}

Assuming a beam quite well approximated by a Gaussian shape, as in
the case of {\sc Planck} beams, the integral of the antenna
pattern is given by $\int_{4\pi} J d\Omega \simeq (\pi/4{\rm ln}2)
({\rm FWHM/rad})^2$. In this work we deal with sources with weak
or moderate flux, i.e. essentially detectable only when they fall
within a pixel with a side $\sim$~FWHM about the beam center. For
a first order evaluation of the source flux sensitivity estimate
we can then neglect the information contained in the time ordered
data outside the pixel with a side $\sim$~FWHM about the beam
center, assume the {\sc Planck} sensitivity per pixel of side
$\sim$~FWHM, and approximate the beam response as unit within a
pixel with a side $\simeq$~FWHM  about the beam center and null
outside in the relationship between the antenna temperature and
the flux reported in the following, i.e. $J / \int_{4\pi} J
d\Omega \simeq 1/({\rm FWHM/rad})^2$ within a pixel with a side
$\simeq$~FWHM  about the beam center and $J / \int_{4\pi} J
d\Omega = 0$ outside. \\
In this approximation, the relationship
between the flux fluctuation and the antenna temperature
fluctuation induced by a discrete source on a squared pixel with
side $\Delta \theta$~(~$\simeq$~FWHM) is:

\begin{equation}
{\delta B_{\nu} /{\rm Jy}
\over [\delta T_{ant} /{\rm mK}] }
\simeq  30.7 [\nu/{\rm GHz}]^2
[\Delta \theta / {\rm rad}]^2 \, ,
\end{equation}

\noindent and the same formula holds for ratio between the rms
noise of the observed flux fluctuation, $\Delta \delta B_{\nu}$,
and the rms noise of the observed antenna temperature fluctuation,
$\Delta \delta T_{ant}$, on a pixel of side $\simeq$~FWHM. This
approximation is used to translate the HFI sensitivities quoted by
\citeasnoun{Puget1998} in terms of total flux fluctuation to the
antenna temperature fluctuation sensitivities reported in Table 1.
More detailed analyses, able to take into account the complexity
of beam shapes, the pointing uncertainty and other relevant
systematical effects will be considered in future technical works
and included in the data analysis pipeline.\\
We will derive the {\sc Planck} sensitivity at the different
channels, relevant for the study of moving objects from the {\sc
Planck} time ordered data, by rescaling the sensitivities of the
final {\sc Planck} maps. At $\nu\lsim100$~GHz it would be possible
to exploit the microwave maps derived from MAP data and cleaned at
the level of MAP sensitivity from the effect of source flux
variations and moving objects. In this case the $1\sigma$
sensitivity levels, relevant for the following discussions
considering the MAP data, will be the sum in quadrature of the
{\sc Planck} sensitivities, appropriately rescaled, and of the
sensitivities of the final MAP data. These values will be also the
sensitivities to be used to detect and to study the moving objects
before producing the {\sc Planck} maps. On the other hand, after
the {\sc Planck} data analysis and the production of the {\sc
Planck} cleaned maps at all frequencies, we may be able to take
advantage from the knowledge of the sky fluctuation in the
positions of the considered moving objects. The sensitivity will
be close to the final one of the {\sc Planck} maps multiplied by
$\sqrt{2}$, i.e. about a factor 2 better than the MAP final
sensitivity in the common frequency range, because a moving object
will not have the same position in the following {\sc Planck} sky coverage.\\

\subsection{{\sc Planck} sensitivity for moving sources}

The {\sc Planck} sensitivity relevant for the study
of moving sources has to be properly
rescaled with respect to that reported in Table 1
to include several factors of sensitivity degradation.\\
A first sensitivity degradation factor, $\sim \sqrt{2}$, derives
by considering only a single sky coverage and provides a lower
limit to the sensitivity utilized in this work (we neglect in the
present estimates the small increase of this degradation factor
with respect to the value $\sqrt{2}$ introduced by the limited sky
regions observed during the mission). The estimate of the {\sc
Planck} best sensitivity for moving sources assumes to be possible
to properly use and to average the information from all the
receivers of a specific frequency channel. Of course, a realistic
situation may be less favorable, because of the motion of the considered object.\\
The LFI beams for each channel are located in a ring with a radius
of about 4$^\circ$, on the {\sc Planck} telescope field of view,
around the line of sight. The HFI beams are located closer to the
center and they may be also at few degrees from the telescope line
of sight. On the {\sc Planck} telescope field of view, two
coordinates U,V are typically used to identify the beam positions.
The direction in the sky of the coordinate V is parallel to the
scan circle of the {\sc Planck} telescope line of sight, generated
by the spacecraft rotation around its spin axis. The direction in
the sky of the coordinate U, orthogonal to V, is, at least for
simple scanning strategies (i.e., with the spacecraft spin axis
always parallel to the antisolar direction),
parallel to the spin axis shift direction during the mission.\\
Given the spread of the {\sc Planck} LFI beams on the telescope
field of view, another degradation factor, $\sqrt{N_{ric,sky}}$,
where $N_{ric,sky}$, ranging from 2 to 16, is the number of
radiometers per channel pointing at the same sky position at the
same time, has to be taken into account.  In column 3 of Table 1
we report the number of LFI radiometers (all intrinsically
sensitive to the polarization) and of the unpolarized and
polarized HFI bolometers. In the case of the LFI channels the
total temperature measurement is given by adding the signals from
the two radiometers coupled to a specific feed horn (or beam);
therefore, $N_{ric,sky} = N_{rad}/2 = N_{feed}$. In all the cases,
except for one feed at 44~GHz, a pair of LFI feeds look at the
same scan circle in the sky, being located in the focal plane unit
to follow the sky scan direction as the spacecraft spins (each LFI
beam, except one at 44~GHz, is symmetrically located to another
LFI beam with respect to the axis V=0, i.e. they have the same U
coordinate).  The factor, $\sqrt{N_{ric,sky}}$, reported in Table
2 at 44 GHz, provides a pessimistic value for our sensitivity
estimates. A different factor, for most of the cases, and reported
in Table 2 for the 30, 70 and 100 GHz channels, is
$\sqrt{N_{ric,sky}/2}$.\\
\begin{table}
\caption{Instrumental noise estimates for moving sources. The
values reported in the columns 2 and 3 refer to the {\sc Planck}
sensitivity for moving sources at middle ecliptic latitudes. In
the case of low ecliptic latitude objects, we report also the $1
\sigma$ overall confusion noise including CMB and astrophysical
fluctuations, neglecting the information from future space
missions (column 4), by considering the information provided by
the MAP survey (column 5), and, finally, the sensitivity
appropriate to a {\sc Planck} single sky coverage (see also the
text). }

 \begin{center}
  \renewcommand{\arraystretch}{0.5}
\begin{tabular}{cccccccccc}
& & \\
\hline
& & \\
$\nu_{\rm eff}$/GHz & $\sigma_{\rm noise}$/$\mu$K & $\sigma_{\rm
noise}$/mJy & $\sigma_{\rm overall}$/mJy &
$\sigma_{\mathrm{after\; MAP}}$/mJy & $\sigma_{\mathrm{after \;
Planck}}$/mJy
\\
& & \\
\hline
& & \\
30  & 7$-$26   & 19$-$70 & 272$-$283 & 53$-$94 & 29$-$83 \\
& & \\
44  & 11$-$76  & 29$-$153 & 249$-$303 & 77$-$189 &
44$-$178
\\
& & \\
70  & 15$-$62  & 40$-$165 & 228$-$293 & 112$-$215 &
61$-$194
\\
& & \\
100 & 18$-$102 & 47$-$266 & 201$-$362 & 138$-$331 &
72$-$309
\\
& & \\
\hline
& & \\
100 & 4.1$-$14.7   & 9.5$-$34 & 193$-$197 & 127$-$133
&
16$-$41 \\
& & \\
143 & 4.8$-$14.9   & 16.3$-$51 & 151$-$161 & 151$-$161
&
25$-$61 \\
& & \\
217 & 4.4$-$11.3   & 16.3$-$42 & 86$-$97 & 86$-$97 &
25$-$51
\\
& & \\
353 & 3.4$-$8.3   & 27.4$-$67 & 49$-$86 & 49$-$86 &
42$-$82
\\
& & \\
545 & 2.8$-$6.9   & 53.7$-$132 & 108$-$175 & 108$-$175
&
82$-$161 \\
& & \\
857 & 1.3$-$3.1    & 60.8$-$149 & 197$-$252 &
197$-$252 &
93$-$182 \\
& & \\
\hline
\end{tabular}
\end{center}
\end{table}
\citeasnoun{Puget1998} obtained the sensitivity on temperature
fluctuation measurements by combining both unpolarized and
polarized detectors for the HFI channels at 143 and 217 GHz. At
545 GHz only polarized detectors will operate, whereas at 100, 353
and 857~GHz only unpolarized detectors are planned.\\
The HFI Focal Plane Unit arrangement is more complex than for LFI
and it will be re-defined in few months (B.~Maffei, private
communication at the {\sc Planck} Systematic Effect Working Group
meeting, ESTEC, June 2001). To translate the sky map temperature
sensitivity of Table 1 to the simple estimate of upper limits for
moving sources quoted in Table 2 we consider that analogously to
the case of the LFI, the HFI feeds at each frequency are aligned
in two or three groups. The feeds of each group, looking at the
same scan circle in the sky, have a very similar value of U. In
the case of the {\sc Planck} HFI, we have to include a further
degradation factor in the range from $\simeq \sqrt{2}$ to $\simeq \sqrt{3}$.
In the sensitivity estimates reported in Table 2
we conservatively use the value $\sqrt{3}$.\\
A last degradation factor $\sim \sqrt{{\rm FWHM} / \Delta
\theta_s}$, where $\Delta \theta_s$ (2.5 arcmin) is the spin axis
shift between two consecutive hours, takes into account the
possibility that a moving object falls out of the beam after a
spin axis repointing. This factor
has to be applied both to LFI and HFI.\\
Taking into account all these sensitivity degradation factors,
the nominal {\sc Planck}
sensitivities can be resumed in the ranges reported in
column 2 and 3 of Table~2;
we provide these sensitivities in terms of antenna temperature and
flux.\\
In practice, we have to keep in mind that the interesting moving
objects are generally located close to the ecliptic plane, where
the {\sc Planck} sensitivity is $\simeq 1.15$ worst than the
averaged sensitivity.  Therefore, the values reported in columns 2
and 3 of Table~2 have to be typically multiplied by $\simeq 1.15$.
By taking this factor into account, we report also in Table~2 the
sum in quadrature of the {\sc Planck} sensitivity for moving
objects as quoted in column 3  and of the astrophysical and
cosmological confusion noise sources of Table~1 (column 4), the
sum in quadrature of the {\sc Planck} sensitivity for moving
objects as quoted in column 3 and of the sensitivity of MAP final
maps (column 5) and, finally, the sum in quadrature of the {\sc
Planck} sensitivity to moving objects as quoted in column 3 and of
the {\sc Planck} sensitivity appropriate to half mission (column
6). These numbers are indicative because of the large variations
of the Galaxy intensity in the sky and the uncertainty of
extragalactic source fluctuations at $\nu \gsim 143$~GHz
(intermediate values are adopted in this work).\\
An important point is represented by the opportunity to combine
the millimetric observations of asteroids performed by {\sc
Planck} with other data obtained in the same period and in
different spectral regions, using ground-based telescopes.\\
The {\sc Planck} asteroid observation at different frequencies
will be obtained within $\sim$ few days, because of the angular
size of {\sc Planck} receiver array and the spin axis repointing
of $1^\circ$ per day. Then we may predict to observe an asteroid
with {\sc Planck} within a week. \\
{\sc Planck} will be injected in its Lissajous orbit around L2
after about 3--4 months from the launch and the {\sc Planck}
nominal scanning strategy will be certainly decided at least many
months before the launch. The effective {\sc Planck} scanning
strategy may be slightly different from the nominal one and will
be accurately reconstructed in few days from the telemetry data.
These requirements will allow to organize an observing campaign,
of the objects detected by {\sc Planck}, on ground-based
telescopes several months in advance.\\
The first rough estimate of the flux for an asteroid observed by
{\sc Planck} will be delivered in a few days, waiting for a longer
period for a more accurate value, at least 6 months corresponding
to the survey duration. This estimate should allow a first
comparison with the ground-based observations.

\section{Threshold detection for asteroids}

In the radio range the mean equilibrium temperature T$_{eq}$ of an asteroid,
averaged over its extent, is given by:

\begin{equation}
{\rm T}_{eq} = f (1-p)^{1/4} r^{-1/2} \, ,
\end{equation}

where $p$ is the bolometric albedo and $r$ the heliocentric
distance. In the nonrotating case $f$ is 329, and for our fast
rotating asteroids (few hours) $f$ is 277. For instance in the
case of 1 Ceres, the mean distance from the Sun is 2.767~AU and
the albedo is 0.06 \cite{Morrison1974},
which give an equilibrium temperature of 165~K.\\
The asteroid equilibrium temperature is related to the asteroid
antenna temperature, both averaged over the asteroid extent, by:

\begin{equation}
T_{\nu,ant,obj}^{\star}= e_{\nu} {\rm T}_{eq} \, ,
\end{equation}

with $e_{\nu}$ the emissivity of the asteroid at the frequency $\nu$.
In this work we need to define a mean value of
${\rm T}_{\nu,ant,obj}^{\star}$ for the main belt
asteroids in order to determine the threshold detection. We will assume
the emissivity close to unity,
since we are considering objects with $T_{eq} \approx 100$~K
and working at millimetric wavelengths.\\
The antenna temperature observed at a specific frequency $\nu$ is
given by the convolution, over the whole solid angle, of the
antenna temperature, of the considered asteroid, with the antenna
beam pattern response (we assume the antenna responses normalized
to the maximum, i.e. $J=1$ at the beam center direction):

\begin{equation}
T_{\nu,ant,obs}(t) = {\int_{4\pi} T_{\nu,ant,obj}({\hat \gamma},t)
~J({\hat \gamma} ,t)~d \Omega \over \int_{4\pi} J({\hat \gamma})~ d \Omega }
\, ;
\end{equation}

in this equation ${\hat \gamma}$ is a unit vector identifying a
generic direction in the sky. As evident from this equation, the
observed antenna temperature depends on the considered observation
time because of the spacecraft rotation and scanning strategy and
the variation of the asteroid properties. Note that, at least in
principle, this equation includes possible deviations from the
uniformity of the asteroid equilibrium temperature on its surface;
since the small angular sizes of the considered objects, they will
be neglected in the following, we will work with quantities
averaged over the asteroid surface. \\
The maximum observed antenna temperature will be obtained when the
antenna beam center points at the asteroid center. By assuming a
bi-dimensional perfectly symmetric Gaussian beam shape and in the
limit of asteroid angular sizes much smaller than the beam FWHM,
the maximum observed antenna temperature can be approximated by
the expression:

\begin{equation}
T_{\nu,ant,obs}^{max} \simeq 4{\rm ln}2 T_{\nu,ant,obj}^{\star} {[R/d]^2
\over ({\rm FWHM/rad})^2} \, ,
\end{equation}

where $T_{\nu,ant,obj}^{\star} \simeq T_{eq}$ ($e_{\nu} \simeq
1$), given by $T_{\nu,ant,obj}^{\star} = \int_{4\pi}
T_{\nu,ant,obj}({\hat \gamma},t) d\Omega / \Omega_{obj}$, is the
average over the object extent of the object antenna temperature
and $\Omega_{obj} = \pi [R/d]^2 $ is the solid angle subtended by
the object, $d$ being its distance from the spacecraft
and $R$ its effective radius at the considered observation time.\\
Assuming $150 {\rm K}$ as mean value of the asteroid antenna
temperature, a geocentric distance range between 2 and 4 AU
($1.5\times10^8$km) and a radius range between 50 and 500 km, we
obtain $R/d$ in the range:

\begin{equation}
R/d \sim 2 \times 10^{-7} - 2 \times 10^{-6} \, .
\end{equation}

These values can be converted in the maximum antenna temperature, for a
specific frequency and beam FWHM, using equation (7):

\begin{equation}
T_{\nu,ant,obs}^{max} \simeq (0.2  - 20) {\rm mK} {1 \over ({\rm
FWHM/arcmin})^2} \, .
\end{equation}

The antenna temperature range for some {\sc Planck} channels is:

\begin{equation}
30 {\rm GHz} \, \, , \, \, {\rm FWHM} \simeq 33'  \, \, \, \, \,
\,
\rightarrow T_{\nu,ant,obs}^{max} \simeq (0.2  - 20) \mu{\rm K}
\end{equation}

\begin{equation}
100 {\rm GHz} \, \, , \, \, {\rm FWHM} \simeq 10'  \, \, \, \,
\, \,
\rightarrow T_{\nu,ant,obs}^{max} \simeq (2 - 200) \mu{\rm K}
\end{equation}

\begin{equation}
217 - 550 {\rm GHz} \, \, , \, \, {\rm FWHM} \simeq 5'  \, \, \,
\, \, \, \rightarrow T_{\nu,ant,obs}^{max} \simeq (10 - 1000)
\mu{\rm K} \, .
\end{equation}

Comparing these values with the noise reported in table 2 we can
see that, mainly at the HFI frequencies, there are good chances to
observe asteroids and for higher $R/d$ even with the LFI channels.\\
In the numerical model we consider only asteroids with $R/d$
larger than $10^{-7}$ (a value referred as ``threshold'' in the
following), since for smaller $R/d$ the maximum signal is less
than about 1~$\sigma_{\rm noise}$ quoted in Table~2 also for the
most favorable channels, and there is then no realistic chance to
derive information on their properties\footnote{It can be noted
that, since the CMB fluctuations dominate the sky confusion noise
between $\sim 50$ and $\sim 250$ GHz, combining more frequency
channels decreases the $1\sigma$ overall confusion noise quoted
for each frequency channel in the last column of Table 2.} (on the
contrary, values of $R/d$ smaller than this threshold have to be
considered in the cleaning of {\sc Planck} data streams from
asteroid contamination).

\section{Identification of target asteroids in the Planck field of view}

\subsection{The numerical algorithm}

To estimate the number of asteroids that will be seen by the LFI
and HFI channels during the {\sc Planck} survey and the precise
dates of observation we have developed a numerical code that
calculates the orbits of all large asteroids and checks if they
fall within the resolution element of a beam. The initial orbital
elements of all asteroids larger than 50 km are taken from a
datafile provided by \citeasnoun{Bowell2000} while the orbits of
the nine planets are derived from the JPL ephemerides at the
corresponding epoch. The diameters are known for most asteroids
larger than 50 km and are reported in Bowell's file. For those
asteroids whose diameter is not reported in Bowell's file, we used
the value given in the file of 12,487 asteroids adopted by
\citeasnoun{Zappala1995} for asteroid family search. Whenever the
albedo is not directly known from observations, an average value,
typical of that particular zone of the asteroid belt, is used to
derive the diameter of the body. To compute the orbits of the
asteroids and planets we have adopted the RA15 version of the
numerical integrator RADAU \cite{Everhart1985} with the highest
value of the precision parameter. \\
At the end of every timestep, we compute the position of each
asteroid respect to the Earth since the {\sc Planck} surveyor will
orbit together with the Earth at its L2 Lagrangian point. Since we
know from the mission specifications the direction of each beam,
we can easily calculate the angle between the asteroid and the
antisolar direction and the center of the beam of each LFI and HFI
channel. A possible future refining will be to compute at each
epoch the position of the spacecraft and compute directly the
angle of the asteroid with respect to the spacecraft. We fixed the
timestep of the numerical integration to 1 hr, that is the time
required by all the channels to complete a 360$^{\circ}$ scan.
Thanks to this choice we can directly compare the position of the
asteroid with all the channel beam directions during the scan at
once. A detection is signalled whenever the angle between the
asteroid and the antisolar direction is within the direction of
the beam divided $+/-$ the FWHM/2.\\
The input data for the code, apart from the orbital parameters,
are the angles that define the inclination of each beam respect to
the antisolar direction. The code begins the comparison between
the position of the asteroids and that of the beam at the nominal
start of the mission and it continues for two years. The code at
the end of the run will give the list of those asteroids that will
be observed by each horn, with the relative date of observation
and $R/d$ ratio. The output include also a list, for each channel,
of the dates when an asteroid will be present in the field of view
of the beam during the scan.

\subsection{Results}

At the end of the simulation 397 asteroids were observed during
the nominal length of the mission in at least one of the horns of
{\sc Planck} with a $R/d$ high enough for individual detection.
For a significant fraction of them (in the range from $\sim 50$ to $100$
objects) the good S/N
ratio will allow to determine the antenna temperature and
information on the physical properties of the asteroid's regolith.
The determination of the antenna temperature at different
frequencies may be useful to derive models for the surface layers
and establish some bounds on the depth of the surface layer and
the
dielectric properties of the material.\\
The times (i.e. the different consecutive scan circles) an
asteroid is consecutively observed by the same horn during a
single passage ranges from 3 to 6 depending on the FWHM of the
channel. It means that on average an asteroid passage lasts about
3--6 hours. In some cases more horns, belonging to the same
frequency channel, may be crossed within some hours allowing to
determine a radio light--curve. In other cases, horns of different
frequencies may be crossed allowing to get information on the
thermal emission of the asteroid. From a statistical point of
view, asteroids will be observed by each channel of both LFI and
HFI in 30\% of the total scan circles performed in the mission.
This datum must be taken into account in the extraction of the
Galactic and extragalactic
radio sources from {\sc Planck} time ordinary data.\\
\begin{figure}
\caption{Number of observations of asteroids by {\sc Planck},
during the whole mission at different sensitivity levels ($\times
10^{-7}$). In the bottom plot at high signal intensity only the 5
largest asteroids appear in {\sc Planck's} horns. The increasing
number of small asteroids that can be observed at lower
sensitivity explains the growing amount of detections. }
\begin{center}
\includegraphics[width=13cm]{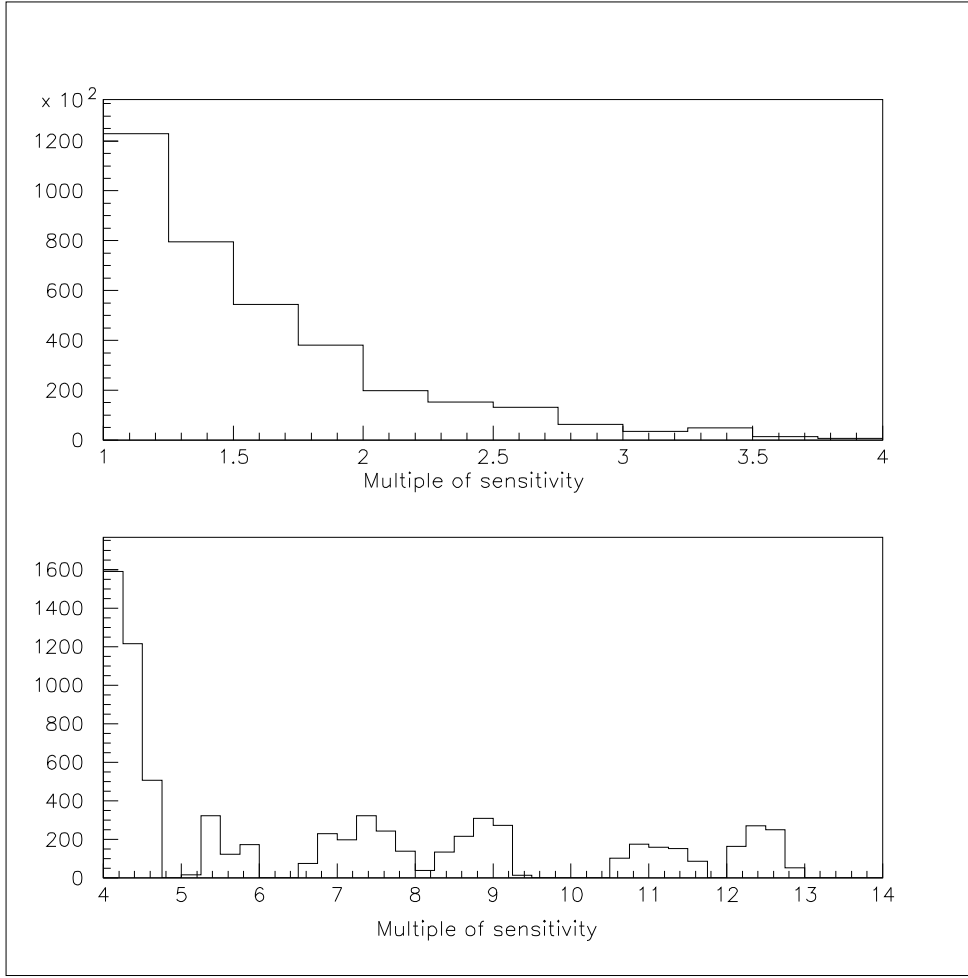}
\end{center}
\label{}
\end{figure}
In Fig.2 the histograms shows the intensity of the signal at
detection as multiple of the minimum detection threshold $R/d \sim
10^{-7}$ for all the observed asteroids on all channels. For large
asteroids the radio signal as detected by the {\sc Planck} horns
reaches values as large as 12 times the detection threshold. In
the plot at the bottom of Fig.~2 we may notice the signals from
the 5 largest asteroids and, so on at smaller $R/d$ ratio, from
all the other visible objects. Of course, the number of detected
asteroids increases drastically if the sensitivity threshold is
reduced since the asteroid size distribution is a power law.  In
Table 3 we report the cumulative and differential number of
asteroids detectable at different levels of the $R/d$ ratio.
\begin{table}
\caption{Cumulative and differential number of asteroids may be
detected at different $R/d$ values.}
 \begin{center}
  \renewcommand{\arraystretch}{0.5}
\begin{tabular}{ccc}
& & \\
\hline
& & \\
$R/d$  & Differential number & Cumulative number \\
\hline
& & \\
1--2 $\times 10^{-7}$ & 299 & 397 \\
2--3 $\times 10^{-7}$ &  76 & 98 \\
3--4 $\times 10^{-7}$ &  15 & 22 \\
4--5 $\times 10^{-7}$ &   4 & 7 \\
5--6 $\times 10^{-7}$ &   0 & 3 \\
6--7 $\times 10^{-7}$ &   0 & 3 \\
7--8 $\times 10^{-7}$ &   0 & 3 \\
8--9 $\times 10^{-7}$ &   1 & 3 \\
9--10 $\times 10^{-7}$&   1 & 2 \\
1--2 $\times 10^{-6}$ &   0 & 1 \\
2--3 $\times 10^{-6}$ &   0 & 1 \\
3--4 $\times 10^{-6}$ &   1 & 1 \\
& & \\
\hline
\end{tabular}
\end{center}
\end{table}

\section{Conclusions}

The {\sc Planck} ESA mission will perform a high-angular
resolution mapping of the microwave sky over a wide range of
frequencies, from 30 to 900 GHz. During the two years planned for
the mission the surveyor will frame a large number of main-belt
asteroids. In this paper we investigate in detail for the first
time the possibility for {\sc Planck} to detect main-belt
asteroids. The main parameter in the asteroid detection process by
the {\sc Planck} horns is the $R/d$\ ratio. The instrumental
sensitivity to such objects has then been defined as the minimum
$R/d$\ value required to safely observe them. This value is given
by the {\sc Planck} sensitivity at different frequencies rescaled
to take into account the integration time and the confusion noise
due to the background. An accurate evaluation of the noises and
the sensitivity degradation, related to the high proper motion of
the asteroids, yielded a minimum $R/d$ ratio
of $(R/d)_{\mathrm{min}} \sim 10^{-7}$.\\
A numerical simulation of a two-years mission has been performed
in order to estimate the number of objects whose $(R/d)$\ is
greater than about  $10^{-7}$\ and that will then be observed by
{\sc Planck}. The simulation uses updated catalogs for the orbital
elements of the main-belt asteroids and their diameters. The orbit
of each asteroid has been integrated with a very short timestep to
compute the relative position of the body with respect to the {\sc
Planck} horns and the value of $(R/d)$. An accurate mission
simulation has been obtained. Up to 397 asteroids are expected to
be detected in the various {\sc Planck} channels (Table 3).
Detectable asteroids will appear in
about 30\% of the total sky circles scanned by the mission. \\
The previous survey of Main Belt Asteroids was performed by the
Infrared Astronomical Satellite (IRAS) on 1983 in four wavelength
bands centered near 12, 25, 60 and 100$\mu$m, much lower than the
minimum wavelength observed by {\sc Planck} of 350$\mu$m,
corresponding to the highest frequency of 857 GHz. It surveyed
approximately 96\% of the sky and 2228 different multiply observed
asteroids were associated to IRAS sources \cite{Tedesco2002},
providing a good estimate of diameter and albedo for most of
them.\\
IRAS's 12$\mu$m limiting sensitivity, for S/N=3, was about 150 mJy
\cite{Tedesco2002b}, and assuming a Main Belt Asteroid as a black
body at 150 K this flux can be translated to about 56 mJy at
350$\mu$m (857 GHz), well below the {\sc Planck} noises reported
in Table 2.\\
{\sc Planck} will provide flux measurements for a smaller sample
of asteroids ($\sim 50-100$ objects) compared to IRAS, but at
different wavelengths, almost unexplored for this class of Solar
System objects. This will improve our understanding of the thermal
emission and the related surface properties of asteroids.

 \section{Acknowledgements}

It is a pleasure to thank
M. Bersanelli, C.R. Butler, K.~Ganga,
D. Maino, N. Mandolesi, F. Pasian, M. Sandri,
F. Villa for useful discussions on
{\sc Planck} performances and simulations, and L. Danese,
G. De Zotti, L. Terenzi,
L. Toffolatti, N. Vittorio for helpful conversations on astrophysical
and cosmological confusion noise. We thank the LFI Consortium
for having promptly provided us with the updated LFI beam
positions and sensitivities.

\newpage

\end{document}